# On some unresolved issues in aggregate level comparative research evaluation

Paul Donner, donner@dzhw.eu, ORCID: 0000-0001-5737-8483, Deutsches Zentrum für Hochschul- und Wissenschaftsforschung, Department 2 Research Systems and Science Dynamics, Berlin, Germany

*Abstract*

This short conceptual paper recalls to the attention of the scientometrics community some long-standing open methodological problems and proposes new directions that might be taken in renewed engagements with them. These problems are the differential dependency of the global reference score on the included units' observed scores and the dilution of impact measurement accuracy due to collaboration, which causes a global convergence in country citation impact due to international collaborative research. Advantages and disadvantages of existing and new solutions are discussed.

## Introduction

Actors such as countries, sub-national states, supranational bodies, state-funded and privately funded foundations all spend vast sums of money on scientific research. In consequence, they are very interested in the performance of the research which they have funded. Bibliometrics, as a method primarily suited to the analysis of large-scale aggregates of research outputs, offers solutions to this demand for performance assessment. These solutions are circumscribed in their capabilities and remain yet distant from the defining, but elusive, methodological challenge of research assessment – the measurement of the progress of knowledge. Bibliometrics must make some unrealistic assumptions, such as the uniformity and commensurability of publications and citations, in order to have basic units of measurement to work with. These limitations notwithstanding, bibliometrics is evidently useful for achieving insights about science as a global socio-cognitive endeavour that cannot otherwise be attained.

However, within bibliometrics as a research community, both the rather simple assessment of a research producing unit in isolation -- counting of publications and their citations to analyze a unit's temporal development -- and the more sophisticated and demanding comparative analysis of units, are far from methodologically settled.

In this conceptual essay I review some rather long-standing unresolved issues in the methodology of bibliometric impact evaluation, that is, assessment of performance in terms of intra-scientific influence and reception by means of citation analysis. They concern the technical side of appropriate and accurate calculation of indicators, rather then more fundamental issues of theoretically informed meaningfulness of citation methods for such purposes.

# Inclusion of target unit in reference group

In the following discussion I will refer to countries for the sake of simplicity, but the arguments may apply also to units of analysis other than countries, whenever aggregate values are needed. This could be the set of publications of individual authors, working groups, departments, organizations, sectors, regions, or multinational bodies such as the EU. However, the substantive issues are most conspicuous and easy to demonstrate for the example of countries.

Suppose we want to compare countries with respect to their scientific impact. A standard approach to comparing countries' (or other units') citation impact, usually understood as an indicator of scientific influence, utility, or a moderately valid indicator of scientific quality (Aksnes, Piro, & Fossum, 2023), is to relate observed citation counts to a global reference score within a defined field of science. For example, the average citation impact values for countries are divided by the world average citation impact value and compared afterwards to make statements such as some country's citation impact 'is above the world average'. One advantage of this approach is that values for different countries are first related to the same reference value and compared to one another only after this operation, which ostensibly has established a common standard.

One little discussed disadvantage of this approach is what may be called the differential dependency of the global reference score on the individual countries' observed scores, which is illustrated in the following example. The problem is especially manifest, but in no way confined to, countries with a relatively large research output.

Suppose that some country *X* accounts for a large share of all global research publications in a field of research. The world reference citation impact score will then be largely determined by *X*'s citation impact. If a relative citation impact score is calculated from *X*'s observed impact and the world reference value, the result will necessarily be strongly influenced by *X*'s large contribution to the world reference value. The relative score relates *X*'s citation impact mainly to its own citation impact.

It makes little sense in a country-comparative scenario to compare any country largely to itself. It is easy to see that the more quantitatively dominant this high-output country is in terms of publication numbers, the less its own impact can possibly deviate from the world reference value. Even if a quantitatively dominant country has exceptionally poor or outstanding citation performance compared to every other country, this difference will be obscured as the world reference value is inevitably similar to the country's own value, the latter being its major contributing component.

The key point is that the highly productive country *X* may have poor or outstanding citation impact in absolute terms, but it will appear to have average impact as soon as conventional relative indicators are calculated.

The general insight to be drawn from this argument is that in this standard approach, because the global reference citation impact value is influenced by countries' individual citation impact scores proportionally to their publication output, countries are compared to a global reference value that is more or less influenced by their own citation impact score. It would thus appear to be a distinct disadvantage of this approach that relative impact scores become less comparatively meaningful with increasing publication production of a country.

Comparing a thing to itself (at the same moment in time) is meaningless. Comparing any thing to some other thing that is nonetheless partially itself becomes less meaningful with increasing partial identity of the compared entities. Or to put it differently, to the extent that one approaches identity, one departs from meaningful comparison.

We are not the first to remark on this conondrum, the following clear expression of it can be found on p. 155 of Schubert, Glänzel, & Braun (1988):

> *It is to be stressed again that the relative indicators used in this study are equally well suited to cross-country and cross-field comparisons provided that the country in question represents only a minor fraction of the publication output and citation impact in the given field. This is usually the case if small or medium-size countries are considered, which contribute only slightly to the world averages used as reference standards in building the relative indicators. On the other hand, the same relative indicators should be used in assessing the major countries ... only with great caution. It may well happen that the publication output and citation impact of one or another research field is dominated by one of these countries ... This will result then automatically in a relative indicator value very close to unity.*

Van Raan (2016, p. 1929) has pointed out the same issue:

> *We notice that for very large size P the value of CPP/FCSm must go asymptotically to 1, because the largest possible P would be all publications worldwide in a specific field, and therefore by definition the field-normalized indicator value has to be 1.*

The problem seems to have been largely neglected since and no consensus has been achieved on how to handle this issue, and the procedure described above remains state of the art.

There are some sensible options for alternative methodological approaches. An obvious and straightforward remedy to this problem would be to compare any country's observed citation impact score to the reference score for the rest of the world, that is, all other countries, excluding the target country itself. Each country's citation score would then be set in relation to a different, country-specific world reference score, a 'rest-of-the-world' score – no longer an expected score, but a contrast score. This surely eliminates the 'contamination' of the reference score by the country's own score. It has the probably undesirable property that in this approach there is no longer a common identical reference value for all countries – the rest of the world is different for each country. On the other hand, it might be considered a particularly overt and thus revealing juxtaposition of the tension between cooperation and competition, which is always present to some degree in the background of such comparisons.

A first example application of this alternative approach –not for countries but for the level of journals– is the Field Normalized Citation Success Index by Tong et al. (2023), which is defined as the probability of the citation count of a journal's paper being greater than the citation count of random comparable paper (same topic, publication year, document type) in other journals. Such probability scores (cf. Crespo et al., 2012), whether defined as excluding the publications of the analyzed unit from the reference set or not, have the drawback that they do not communicate how much more or less frequently the unit’s publications are cited compared to the reference value. Another example is the external rhythm framework of Liang & Rousseau (2026), in which the performance of a unit as a member of a set of units is compared to the performance of the units forming the complement in that set.

Another option would be to calculate the world reference score with equal weights for each country's citation impact contribution. This would also eliminate the differential dependency. Each country's impact would still be compared partially to its own impact but the size of this 'contamination' would be exactly equal for each country and the feature of a common reference value is retained.

A third option would be simply not to calculate any world reference score, but to compare observed country scores without relating them to any world reference value. This would only work within fields as it dispenses with the field normalization achieved by creating relative indicators within fields.

# ‘Dilution’ of scientometric impact measurement accuracy by increasing international collaboration

Suppose one would like to assess the typical citation impact of an author. Authors may investigate and publish by themselves or in collaboration with colleagues. As a manifestation of an author's work, the author’s single-authored publications are a better source of information of the author’s citation impact, understood as a proxy of an intrinsic capability for research quality, compared to co-authored publications. The impact of the co-authored publications, in constrast, is a reflection of the combined input of the co-authors and as such a less accurate reflection of any single co-author's typical impact. This reasoning naturally scales up to collaborative work by multiple research groups, organizations, and countries. The aggregate citation impact of internationally co-authored publications of a country of interest will be affected by the intrinsic national citation impact capacities of the participating countries, ‘diluting’ the value of such publications for expressing the impact of any single participant country. As Persson, Glänzel, & Danell (2004) observe:

> *at the level of countries an increased level of collaboration could lead to diminishing differences among the countries in citation impact – and thus in a sort of convergence.*

This can perhaps to some extent be counteracted by weighting citation scores of internationally co-authored publications differently by the estimated extent of participation of countries derived from co-authorship credit allocation schemes, of which there are many (Gauffriau, 2021).

A consequence of the ever-increasing level of international research cooperation is that country impact scores, conventionally measured, have in fact been globally converging. Not because the intrinsic research capacity of countries worldwide has been levelled, but because every country’s research output is increasingly part of the global cooperative output contributing to the scores of every country to the degree that they are engaging in international cooperation (Adams & Szomszor, 2022, p. 726). Persson et al. (2004) found citation impact convergence only for the more cited countries, comparing standard deviations of observed citation rates of 1980 and 1998 international papers. Adams and Szomszor (2024) find that during 1981-2018, for most countries, domestic and international impact, analyzed separately, remain largely stable. Yet, net national scores change and converge as research becomes increasingly internationally collaborative.

This ‘dilution’ of measurement precision and the correlation between citation impact and number of involved authors, institutions, and countries (e.g. Katz and Hicks, 1997) can combine to cause dramatically distorting effects on national indicator values, particularly if arithmetic means are used. For example, Aksnes & Sivertsen (2004) report:

*In one instance in Physics, we found that the occurrence of one highly cited paper influenced the national citation indicator to a high degree without having any relevance as an indication of 'national performance'. A highly cited CERN-article accounts for 23% of all citations to Norwegian articles in Physics in a particular year, while this share fluctuates between 2% and 8% in all other years. This article has an extremely large number of authors from many countries. One institutional address in Norway significantly affects the national citation indicator for Physics that year.*

It would have been an attractive solution to ask the Norwegian co-authors to this publication for an estimate of their relative contribution and then use this value as a weigthing factor or correction factor for the citation count value. As such data collection does not realistically scale, we have to make do with a surrogate method of estimating the contribution shares to co-authored publications with appropriately validated co-authorship credit allocation schemes (counting methods). Another simple option to deal with this is to focus analysis on those publications in which the units under analysis have not collaborated with each other. For example, for a comparison of EU and US output, one defines the EU publication group as all those publications with only within-EU addresses, the US group as those with only US addresses, and one can further include EU-US collaborative publications for additional context (Rodríguez-Navarro & Narin, 2018).

# Proposal for a new method

An additional option of handling the issues arising from different sizes of output and ubiquitous collaboration can be proposed. The idea is to compare the score of each country's single-country papers to the common reference score of all papers produced in any international collaboration. This contrasts national to international research and retains the advantage of one common reference group and reference score for all countries without being afflicted by differential dependency of the reference score, which biases large countries' score towards the reference value. In this scenario, abstracted from the level of countries, the basic principle of comparison becomes the distinct individual impact of each unit, including internal (countries: domestic) collaborative impact, to the average impact of collaboratively (countries: internationally) produced science.

This perspective is unfamiliar in bibliometrics, but it is worth a serious consideration and further study. It is obviously hardly applicable on the level of authors.

In this mode, clearly the reference value is still affected more by the values of larger units, but, in contrast to the standard approach, it is no longer the same publications which contribute to both the target unit's individual value and the global aggregate value, because these two publication sets are always distinct. One is then no longer comparing a large unit's value mainly to itself, but two analytically dinstinguishable parts of the unit's output, only one of which, the one entering the calculation of the global reference value, is furthermore bound up with those of compared units, while the other is as directly reflecting the unique 'undiluted' research performance of the target unit as can be achieved.

The following thought-experimental argument provides further support. Suppose one would like to compare the impact of two comparable units only to each other without taking into account further global reference units at all. If these units, say, countries in a specific discipline, produced some publications collaboratively, it is justified to exclude them from the comparative analysis or analyze them seperately as a distinct group, on the grounds that they do not reflect well the performance of any single one of the two units, but their combined performance, which ultimately cannot unambiguously be attributed partially to the two units. Their inclusion in the calculation of any bibliometric scores on both sides of the comparison would be logically redundant in this isolated comparison scenario.

We may thus have arrived at a series of increasingly higher order approximation models of research units' performance, which are successively more accurate with respect to reflecting their intrinsic individual research impact capacity, which is conceptually and practially different from their collaborative and their 'total' (individual + collaborative) research impact capacity. The first order approximation model is whole counting of publications and citations, where the reference group includes the target unit. At the level of second order approximation models, several options exist:

- The calculation of 'rest-of-the-world' reference scores particular to each unit.
- Equal weighting of each unit's contribution to the global reference score.
- Dispensing altogether with a global reference score when possible.
- Calculating individual unit scores of single-unit publications and comparing to a global reference value of publications by multiple units.

Some of these models may in turn be improved to be a third order approximation for application at the person level by abandoning the unrealistic assumption of equal contribution of all co-authors to co-authored papers and using the most accurate available credit allocation method as weighting factors.

# Indicator target quantities

At any level of aggregation, indicators can target different quantities, or concepts. Important concepts for which indicators are used are average citation impact, total knowledge contribution, and top-level knowledge contribution (excellent or breakthrough science). All of these can be seen as properties of the research quality distribution, or profile, of a unit: the distribution's central tendency (which may not exist), its area, and the area of a defined part of its upper tail. The study of the whole citation impact distribution for evaluative purposes has been proposed repeatedly (e.g., Adams et al., 2019; Glänzel & Schubert, 1988; van Leeuwen et al, 2003). The efficiency of knowledge production is also often of great interest and requires additional information on process input, such as personnel quantity and quality and financial input. It seems plausible that the above methodological considerations apply regardless of the targeted concept or for the whole-distribution approach.

# The validation impasse

The foregoing discussion is entirely founded on theoretical-conceptual rational argumentation. No empirical data has been put forth in support of one method or the other. This is no oversight. At the macro level of citation impact methodology for countries, empirical validation with external criterion data is almost inconceivable. We could calculate some example scores for the different approaches and show that they lead to different results, but we would be hard pressed to show which of the different methods produces scores that are closer to the actual true scores, which remain elusive. Acceptable validation of bibliometric methods at lower levels seems more feasible, when one is willing to rely on peer review data as a benchmark. One could then make the argument that a methodological approach that better reflects peer review judgments at author or department level should also better reflect country impact. Nevertheless, we have to concur with Narin's (1976) remark:

> *The fundamental problem in any discussion of the validity of indicators of scientific productivity is the fact that there is no absolute standard of measure of such productivity. The classic scientific approach would be to obtain the best possible set of indicators of quality or productivity from the literature, and validate these indicators through correlations and similar analyses with independent, objective and quantitative measures of scientific productivity and quality. Unfortunately, no such set of independent, objective, quantitative indicators exists.*

Nor has any become available since.

# Conclusion

In this conceptual paper I have revisited two long recognized methodological issues in evaluative bibliometrics which have remained unresolved. I have called these the differential dependency of the global reference score on the included units' observed scores and the dilution of impact measurement accuracy due to collaboration, elsewhere also concetualized as an (artifactual) global indicator score convergence. Proposed solutions' advantages and limitations have been discussed. It should be noted that in many studies, including those designed as input for policy-making processes, these issues are not given any attention at all. In the hope that this necessarily selective survey might reinvigorate methodological experimentation, I have also proposed a novel approach at dealing with the above issues, namely to compare units' scores for their independently produced output with a reference score calculated from all units' collaboratively produced output.